\documentclass[aps,showpacs,prb,twocolumn,amsmath,amssymb,superscriptaddress]{revtex4-1}
\usepackage{bm}
\usepackage{amssymb}
\usepackage{colordvi}
\usepackage{graphicx}
\usepackage{color}
\usepackage{hyperref}

\begin{document}

\title{Dephasing measurements in InGaAs/AlInAs heterostructures: \\
manifestations of spin-orbit and Zeeman interactions}

\author{L. H. Tzarfati}
\affiliation{Raymond and Beverly Sackler School of Physics and Astronomy, Tel Aviv University, Tel Aviv 69978, Israel}
\email{orawohlman@gmail.com }

\author{A. Aharony}
\affiliation{Raymond and Beverly Sackler School of Physics and Astronomy, Tel Aviv University, Tel Aviv 69978, Israel}
\affiliation{Physics Department, Ben Gurion University, Beer Sheva 84105, Israel}
\author{O. Entin-Wohlman}
\affiliation{Raymond and Beverly Sackler School of Physics and Astronomy, Tel Aviv University, Tel Aviv 69978, Israel}
\affiliation{Physics Department, Ben Gurion University, Beer Sheva 84105, Israel}

\author{M. Karpovski}
\affiliation{Raymond and Beverly Sackler School of Physics and Astronomy, Tel Aviv University, Tel Aviv 69978, Israel}
\author{V. Shelukhin}
\affiliation{Raymond and Beverly Sackler School of Physics and Astronomy, Tel Aviv University, Tel Aviv 69978, Israel}
\author{V. Umansky}
\affiliation{Department of Condensed Matter Physics, Weizmann Institute of Science, Rehovot, 76100, Israel}
\author{A. Palevski}
\affiliation{Raymond and Beverly Sackler School of Physics and Astronomy, Tel Aviv University, Tel Aviv 69978, Israel}

\date{\today}

\begin{abstract}

We have measured weak antilocalization effects, universal conductance fluctuations,  and Aharonov-Bohm oscillations in the 
two-dimensional electron gas formed  in  InGaAs/AlInAs heterostructures. This system  possesses strong spin-orbit coupling and a high Land\'{e} factor.
Phase-coherence 
lengths of  2$-$4 $\mu$m at 1.5$-$4.2 K are extracted from the magnetoconductance measurements. 
The analysis of the coherence-sensitive data reveals  that the temperature dependence of the decoherence  rate complies with the dephasing mechanism originating from electron-electron interactions in all three experiments. Distinct beating patterns superimposed on  the Aharonov-Bohm oscillations are observed over  a wide range of magnetic fields,  up to 0.7 Tesla at the relatively high temperature of 1.5 K. The possibility that these beats are due to the interplay between the Aharonov-Bohm phase and the Berry one, different for  electrons of opposite spins in the presence of  strong spin-orbit and Zeeman interactions in ring geometries,  is carefully  investigated. It appears that our data are not explained by this mechanism; rather, a few geometrically-different electronic paths within the ring's width can account for the oscillations' modulations.
\end{abstract}


\maketitle
\section{Introduction}
\label{intro}

The  electronic characteristic scale on which quantum interference can occur in a mesoscopic sample
is the phase-coherence length $L_{\phi}$.
The study 
of decoherence  in quantum-mechanical systems has gained much interest recently,  because $L_{\phi}$  is relevant to  spintronics, i.e., to spin-sensitive devices \cite{Fert,Wolf,Zutic,Bratkovsky} comprising materials
with strong spin-orbit interactions. 
The variation of $L_{\phi}$
with the temperature $T$ serves to indicate the main scattering mechanism which limits phase coherence, be it electron-electron, electron-phonon, or spin-dependent,  scattering processes.  At low temperatures,  though,  
electron-electron
scattering is the dominant  mechanism responsible  for  dephasing. 
Theoretically, the dephasing rate, $1/\tau_{\phi}$, 
due to this scattering  vanishes linearly with $T$ as the temperature decreases towards 
zero,  in agreement with the prediction of Altshuler {\it et al.} \cite{Altshuler} To
determine experimentally the relevant dephasing mechanism and to estimate the coherence length, 
quantum-interference properties, such as weak localization and antilocalization,  \cite{Hikami,Maekawa}
universal conductance 
fluctuations,  \cite{Lee}  and Aharonov-Bohm  oscillations,  \cite{Aharonov,Washburn,Milliken} are 
measured  and analyzed. These  quantum effects have
different dependencies on the coherence length;  their combined
study 
provides a comprehensive picture of the processes leading to decoherence in weakly-disordered nanostructures.

Here we focus on nanostructures in which the electrons are subjected to significant spin-orbit coupling, and
report on studies of weak antilocalization (WAL) effects, universal conductance fluctuations  (UCF), and Aharonov-Bohm (AB) oscillations in the magnetoresistance data of mesoscopic samples of InGaAs/AlInAs. This material is well-known for its strong Rashba-type spin-orbit interaction, \cite{Rashba,Winkler} characterized by the  coupling strength $\alpha_{\rm so}$  of about of $ 10^{-11}$eV m. \cite{Schapres,Nitta} This value corresponds to a spin-orbit energy \cite{Aronov} $\hbar\omega_{\rm so}
=(m^{\ast}v_{\rm F}/\hbar)\alpha_{\rm so}\approx 1.6 $ meV (the Fermi wave vector of our samples is $\approx 1.58\times 10^{6}$ cm$^{-1}$). The Land\'{e} factor of our material is about 15, and hence the Zeeman energy is $\hbar\omega_{\rm Z}\approx 0.87 \times B$ meV, where the magnetic field $B$ is measured in Tesla.

The spin-orbit interaction,  coupling the momentum of the  electron to its spin,   in conjunction with a Zeeman field  gives rise to  Berry phases. \cite{Berry} The simplest illustration of a Berry phase occurs when a spin 1/2 follows adiabatically   
a magnetic field whose direction varies in space. \cite{Engel,Stern}
When that direction returns 
to its initial orientation the spin wave function acquires a geometrical phase factor.
A spatially-inhomogeneous magnetic field can be produced by the joint operation of  spin-orbit coupling and a Zeeman field. \cite{Aronov} Because the Berry phase may modify periodicities related to the Aharonov-Bohm effect, it
has been  proposed that it   can be detected in persistent currents,  
magnetoconductance, and universal conductance fluctuations
of strongly spin-orbit coupled mesoscopic systems. \cite{Stern,Aronov,vanLangen,Loss} Specifically, 
the Berry phase  is expected to manifest itself in additional oscillations superimposed on the conventional Aharonov-Bohm ones, leading to peak-splitting in the power spectrum of those oscillations, \cite{Engel}
i.e., to a beating pattern.
Beating magnetoconductance oscillations have been indeed reported   \cite{Morpurgo,Yau,Lyanda,Grbic,
Meijer}  for AB rings  fabricated in materials with strong spin-orbit interactions  at temperatures below 500 mK. In comparison, our samples show  beating patterns at much more elevated temperatures.

However, one should  exercise caution when adopting the interpretation based on the effect of  Berry phases  for  beating patterns superimposed on Aharonov-Bohm oscillations.
First, the Aharonov-Bohm oscillations 
appear at arbitrarily small magnetic fields, while the effect of the Berry phase 
reaches its full extent 
only in the adiabatic limit,
realized when 
both $\omega_{\rm Z}$ and $\omega_{\rm so}$ are larger \cite{Aronov,Engel,Stern} than 
the frequency of the electron rotation around the ring.
Second, the Berry geometrical phase is restricted to the range $\{0,2\pi\}$, limiting  the corresponding geometrical flux
to the order of one flux quantum,  \cite{Stern} which may make it negligible as compared with the Aharonov-Bohm flux. Third, there can be other causes for the appearance of beating patterns:
a recent experimental study \cite{Ren} carried on 
InGaAs/InAlAs
mesoscopic rings reports on
beating patterns in the magnetoresistance as a function of the magnetic field, measured at temperatures up to 3 K. The authors attribute these patterns to the interplay of a  few,  geometrically-different, closed 
paths that are created in  a finite-width  ring.  \cite{Aharony} We  carry out below a thorough attempt to fit our AB oscillations' data to the theoretical expressions predicting the beating patterns, in particular the expressions given in Ref. \onlinecite{Aronov}.  We find that the theoretical expression for the transmission of a strongly spin-orbit coupled  Aharonov-Bohm ring does show a beating pattern. 
However, it seems to be due to the Zeeman interaction alone; the reason being the confinement of the Berry phase to the range $\{0,2\pi\}$ mentioned above. Our conclusion is that, given the physical parameters of our rings, the beating patterns we observe probably  cannot be attributed to the effects of the Berry phase.

The remaining part  of the paper is organized as follows. Section \ref{preparation} describes the samples' preparation and the measurements techniques. Section \ref{results} includes the results of the measurements of the antilocalization effects (Sec. \ref{WAL}), the universal conductance fluctuations (Sec. \ref{UCF}), and the Aharonov-Bohm oscillations (Sec. \ref{AB}). In each subsection we list the values of the coherence length extracted from the data.   In the last subsection there we combine the results of all measurements to produce the dependence of the dephasing rate in our samples on the temperature (Sec. \ref{rate}), from which we draw the conclusion that it is electron-electron scattering that dephases the interference in our InGaAs/AlInAs heterostructures.
Section \ref{beating} presents our attempts to explain the beating pattern of the AB oscillations displayed in Sec. \ref{AB}. Our conclusions are summarized in Sec. \ref{summary}.

\section{Samples' preparation and measurements}
\label{preparation}

\begin{figure}
\includegraphics[width=6cm]{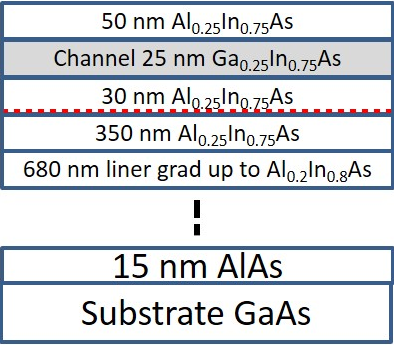}
\caption{
(Color online)  Schematic structure of the sample layers. The dashed (red)  line   in  the spacer layer       is the Si $\delta-$doping. 
}
\label{Fig1}
\end{figure}

Three types of samples were prepared,  all comprising a single basic material. The schematic drawing of the layers in the InGaAs/AlInAs heterostructures used in our studies is given in Fig. \ref{Fig1}. This material was grown by molecular-beam epitaxy,   as described in detail elsewhere. \cite{Hoke,Zeng} The geometrical shape of  above-micron devices  was patterned by a conventional photolithography, while that   of  the nanoscale ones were patterned using e-beam lithography. About 1 micron deep mesa was etched with phosphoric acid (of concentration 1:8) to prevent  as much as possible parasitic  conduction in the structure below the quantum well. Vacuum deposition of a Au-Ge conventional alloy was used to form Ohmic contacts. Electron density of 4.55 $\times 10^{11}$ cm$^{-2}$ and electron mobility of 1.8 $\times 10^{5}$ cm$^{2}$/(V sec) were deduced from  resistivity and  Hall-effect measurements taken at 4.2 K.
These values were calculated for the samples which have a significant contribution of the parallel conduction of low mobility layers below the 2DEG in the quantum well, and therefore are different from the actual values of the mobility and carrier density of electrons in that quantum well.

\begin{figure}
\includegraphics[width=8cm]{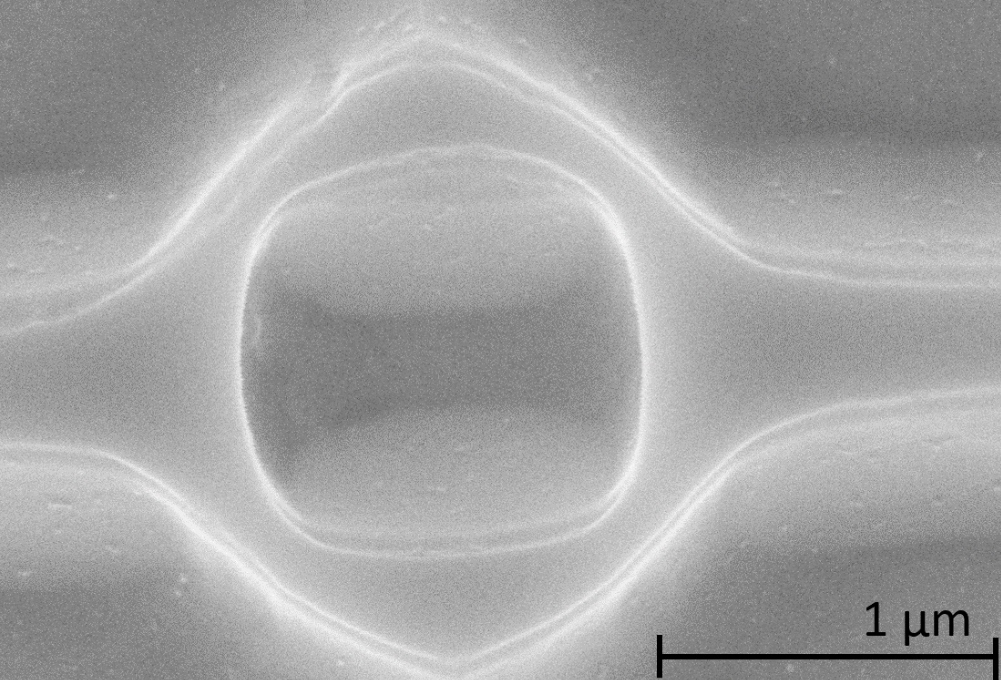}
\caption{
(Color online)  
High-resolution scanning-electron microscope    image of one of the measured Aharonov-Bohm rings. }
\label{Fig2}
\end{figure}

Measuring each of the coherence effects requires samples of different geometry. We have used  a 110 $\mu$m long (i.e., the distance between the voltage probes) and 10 $\mu$m wide Hall bar for the weak-localization studies,  a shorter Hall bar of length 8 $\mu$m  and   width
0.2 $\mu$m for the UCF measurements, and  two identically-prepared  rings (denoted below by ``A" and ``B"),  of  average radius  0.75 $ \mu$m, and  average width  0.2 $\mu$m for the AB measurements,      see  Fig.  \ref{Fig2}.
The resistance was measured by the four-terminal method,  exploiting a low-noise analog lock-in amplifier (EG\&GPR-124A) in  perpendicularly-applied magnetic fields up to 5 Tesla. The measurements were performed in a $^{4}$He cryostat at temperatures in the range of 1.4$-$4.2  K.

\section{Results }
\label{results}

\subsection{Weak antilocalization}
\label{WAL}

\begin{figure}
\includegraphics[width=8.5cm]{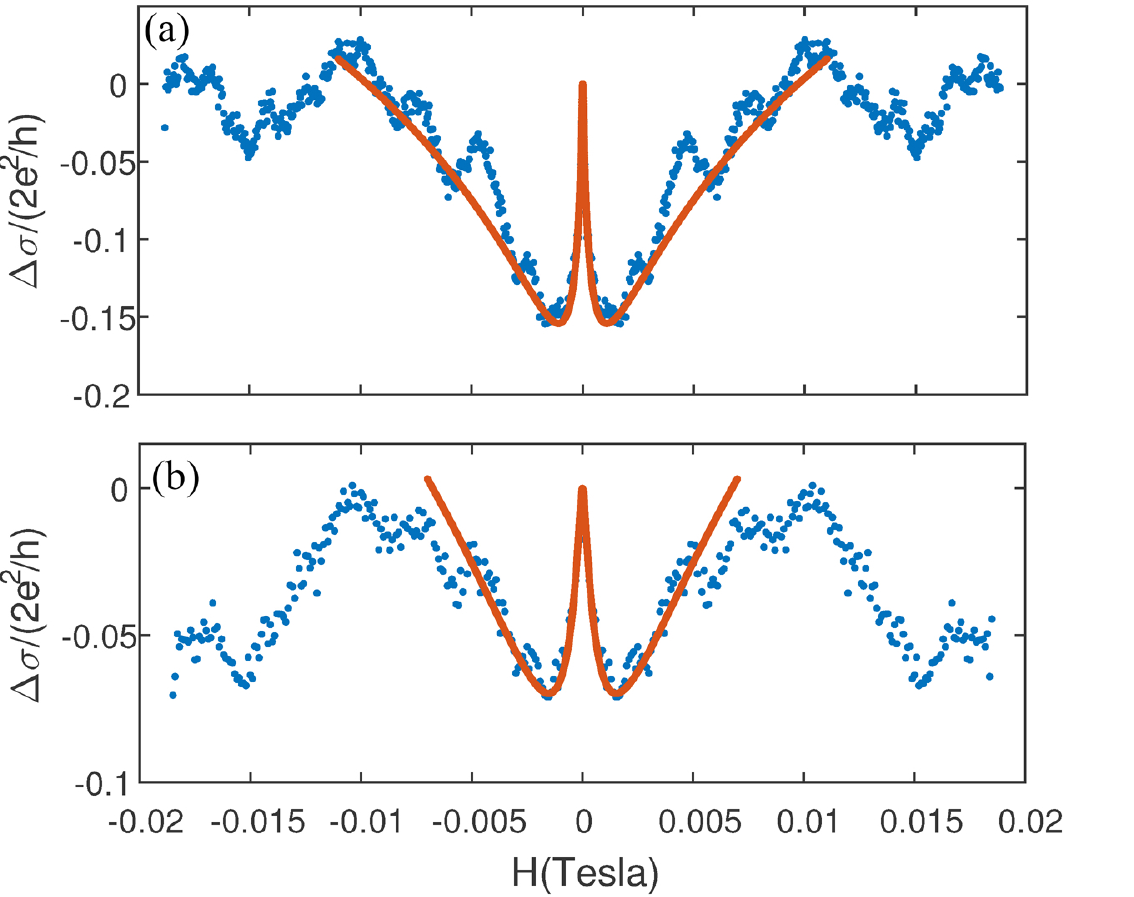}
\caption{(Color online) The magnetoconductivity as a function of a magnetic field normal to the sample plane,  at 1.6 K (a)  and 4.2 K (b), for the WAL sample.  The dotted (blue) lines are the data; the solid (red) curves represent the theoretical magnetoconductivity, calculated from Eq. (\ref{ds}).
 }
\label{Fig3}
\end{figure}

Weak-localization corrections to the average conductivity  arise from  interference between pairs of time-reversed paths that return to their origin.  Application of a magnetic field that destroys  time-reversal symmetry suppresses the interference and thus increases the conductivity. 
Antilocalization appears in  systems in which the electrons are subjected to  (rather strong) spin-orbit coupling. Then,  the interference-induced correction to the conductivity   is reduced, 
because the contribution of time-reversed paths corresponding to wave functions  of opposite spins' projections is negative, while that of the equal  spin-direction time-reversed paths remains positive. 
The reason is that upon following a certain closed path,   the electron's spin is rotated by $\pi$, while for the time-reversed  path with the opposite spin projection  it is rotated by $-\pi$. These two phases add up to give a total rotation of $2\pi$,  leading to a Berry's phase factor of $-1$.
This results in a higher  net conductivity, and
 the positive magnetoconductivity  caused by localization is turned into  a negative one at low magnetic fields.

Measuring the magnetoconductivity as a function of the magnetic field allows for an accurate estimate of the phase-breaking length $L_{\phi}$.   The dotted
curves in Fig. \ref{Fig3} are the
magnetoconductivity  $\Delta \sigma(B)$ 
of  the longer Hall bar as a function of a magnetic field $B$ directed normal to the sample. 
Upon increasing the magnetic-field strength from zero,  one observes a decreasing conductivity originating from the suppression of antilocalization,  followed by an increase due to the destruction of localization. 
Indeed, the  line shapes at small magnetic fields  measured at 1.4 K and 4.2 K,  are nicely fitted to the  curves calculated from the theoretical expression   derived in  Refs. 
\onlinecite{Hikami} and \onlinecite{Maekawa}. As found there, the magnetoconductivity  of a two-dimensional electron gas,  in the presence of a perpendicular magnetic field, is 
\begin{align}
\Delta\sigma(B)&\equiv \sigma(B)-\sigma(0)\nonumber\\
&=-\frac{e^{2}N_{v}\alpha }{2\pi\hbar^{2}}[\Psi (x_{1}^{})-\frac{3}{2}\Psi (x^{}_{2})+\frac{1}{2}\Psi (x^{}_{3})]\ ,
\label{ds}
\end{align}
where
\begin{align}
\Psi(x)=\ln (x)+\psi( [1/2]+[1/x])\ ,
\end{align}
$\psi$ being the digamma function.
In Eq. (\ref{ds}), $N_{v}\alpha$
is the valley degeneracy, and
\begin{align}
x^{}_{1}=\frac{B}{B^{}_{0}+B^{}_{\rm so}}\ ,\ \ x^{}_{2}=\frac{B}{B^{}_{\phi}+\frac{4}{3}B^{}_{\rm so}}\ ,\ \ x^{}_{3}=\frac{B}{B_{\phi}}\ .
\label{x}
\end{align}
These parameters comprise  $B_{\phi}=\hbar/(4eL^{2}_{\phi})$,   the ``phase-coherence" magnetic field, roughly the 
field required 
to destroy phase coherence, $B_{\rm so}=\hbar/(4eL^{2}_{\rm so})$
that represents the spin-orbit coupling, with $L_{\rm so}\approx v_{\rm F}/\omega_{\rm so}$, 
and 
$B_{0}=\hbar/(4e\ell^{2})$, where $\ell$ is the mean-free path.

The comparison of  the data with Eq. (\ref{ds}) has yielded     
 $L_{\rm so}=0.87\pm0.09\ \mu$m for the spin-orbit characteristic length,  
$ L_{\phi}=3.9\pm 0.9\ \mu$m 
at 1.6 K, and 
$ L_{\phi}=1.7\pm 0.3\ \mu$m  at  4.2 K for the phase-coherence length. The relatively large error bars 
do not arise from the fitting procedure; these are due to the scattering of the fitting values obtained for different samples.

As seen in Fig. \ref{Fig3}, 
the curves of  the data-points  deviate from the theoretical ones for magnetic fields exceeding $B=0.01$ Tesla. 
We believe that at these  fields there appear other quantum corrections, e.g., interaction effects,  and contributions arising from the parasitic conductances of the layers below the quantum well.

Equation  (\ref{ds}) derived in Refs. \onlinecite{Hikami} and \onlinecite{Maekawa}
emphasizes the contribution to the conductivity resulting from the  impurity-induced  spin-orbit interaction or from the cubic  (in-the-momentum) Dresselhaus coupling. The theory of Iordanskii {\it et al.} \cite{Iordanskii}
accounts for the linear-in-the-momentum Rashba interaction, which is rather significant in InGaAs. \cite{Knap}
As shown in Ref. \onlinecite{Iordanskii}, this linear interaction adds another characteristic spin-orbit field in addition  to $B_{\rm so}$,   representing  the linear-in-the-momentum Rashba interaction. This additional field is denoted $B_{\rm so}'$.
Indeed, our data can be  also fitted  to  Eq. (13) of Ref. \onlinecite{Iordanskii}; we have found though, that due to the larger number of fitting parameters [as compared to Eq. (\ref{ds})]   multiple sets of  the fitting parameters $B_{\rm so}$ and $B_{\rm so}'$ can  produce the same quality of  fit as obtained for  Eq. (\ref{ds}),  with the same values of $B_{\phi}$  as used in the latter. In order to distinguish between the sets of fitting parameters the range of magnetic fields should be much larger and the quality of the data, limited mostly by universal conductance fluctuation, should be much better. Unfortunately, out data  do not meet these restrictions. As the focus of the present study is on  dephasing mechanism, and since both theories produce 
the same values of $B_{\phi}$,  we present in Fig. \ref{Fig3} the fitting curve of Eq. (\ref{ds}).

Finally we note that for $R_{\square} =220$ $\Omega$ and carrier density of $4\times 10^{11}$ cm$^{-2}$, the mean-free path is about 0.3 $\mu$m, which gives $B_{0}\approx 2\times 10^{-3}$ Tesla.  We have found that $L_{\phi }$  is not very sensitive to the value of $B_{0}$.  

\subsection{Universal conductance fluctuations}
\label{UCF}

Like weak localization and weak antilocalization effects, the universal conductance fluctuations 
of a mesoscopic system result   from
interference of the electronic wave functions corresponding to pairs of  time-reversed paths. As such, these fluctuations are dominated by the phase-coherence length $L_{\phi}$. The UCF are expressed by the ensemble-average autocorrelation function of the dimensionless conductance,   $g=G/(e^{2}/h)$,  \cite{Lee}
\begin{align}
F(\Delta B)=\langle \delta g(B)\delta g(B+\Delta B)\rangle \ , 
\end{align}
where
$\delta g(B)=g(B)-\langle g(B)\rangle$. The angular brackets denote the ensemble average. 
Theoretically, the  average is  over an ensemble of mesoscopic systems   of various impurity configurations; the experiment is carried out on a single sample and the average is accomplished  by 
ramping a magnetic field over the range $\Delta B$ ($\Delta B$ was in the range $10^{-3}-1$ Tesla).   This can generate sample-specific, random-looking but reproducible fluctuations in the conductance.

The
phase-coherence length 
is derived from the magnetic correlation field $B_{c}$, i.e., the field corresponding to the  half width at half height of $F$. This magnetic correlation field is found from the correlation function using the condition
\begin{align}
F(\Delta B=B^{}_{c})=\frac{1}{2}F(\Delta B=0)\ , 
\end{align}
where $
F(\Delta B=0)$ is the
root-mean-square (rms) of the conductance fluctuations, $\Delta g$, 
\begin{align}
\Delta g&=\frac{N^{}_{v}\alpha }{\beta}\Big (\frac{L^{}_{\phi}}{L}\Big )^{3/2}\ ,
\label{Dg}
\end{align}
($L$ is the length of the specimen). \cite{Beenakker} The coefficient $\beta$ represents the effect of the spin-orbit coupling on the magnitude of the fluctuations.
The correlation field is given by
\begin{align}
B^{}_{c}=\frac{h/e}{WL_{\phi}}\ ,
\label{Hc}
\end{align}
where $W$ is the sample's width.

\begin{figure}
\includegraphics[width=8.5cm]{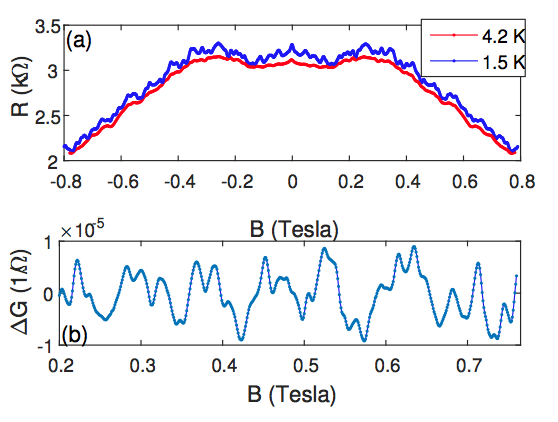}
\caption{(Color online) (a)  The resistance as a function of the magnetic field of a UCF sample at 1.52 K and  at 4.2  K.  (b) The deviation of the magnetoconductance from the average background average. }
\label{Fig4}
\end{figure}

The resistance of the shorter Hall bar, 
measured at 1.52  K  and at  4.2 K, is shown in Fig. \ref{Fig4}(a). The reproducible conductance fluctuations  are displayed  in Fig. \ref{Fig4}(b); the curve there is obtained by  subtracting the slowly-varying background of the average conductance from the measured one.  
Taking $\beta=2$ (corresponding to strong spin-orbit coupling \cite{Beenakker,Chandrasekhar})
in Eq. (\ref{Dg})   yields that the coherence length of our short Hall bar is
$L_{\phi}=3.03\pm 0.8 \ \mu$m at 1.52 K and is 1.56$
\pm 0.8 \ \mu$m at 4.2 K;
Eq. (\ref{Hc}) yields the values 
$L_{\phi}=2.3\pm 1.2 \ \mu$m at 1.52 K and 1.65$
\pm 0.25 \ \mu$m at 4.2 K.

\subsection{The frequency and the amplitude of the Aharonov-Bohm oscillations}
\label{AB}
Perhaps the
most conspicuous manifestation of the Aharonov-Bohm effect \cite{Aharonov} in condensed matter
are the periodic oscillations of the magnetoconductance of a mesoscopic ring as a function of the magnetic flux penetrating it,  whose periodicity  is the flux quantum  $\Phi_{0}=h/e$.
These oscillations are utilized to probe the sensitivity of the electronic wave functions to magnetic fluxes.
Their amplitudes, i.e., their ``visibility" is the hallmark of quantum coherence. 

\begin{figure}
\includegraphics[width=8.5cm]{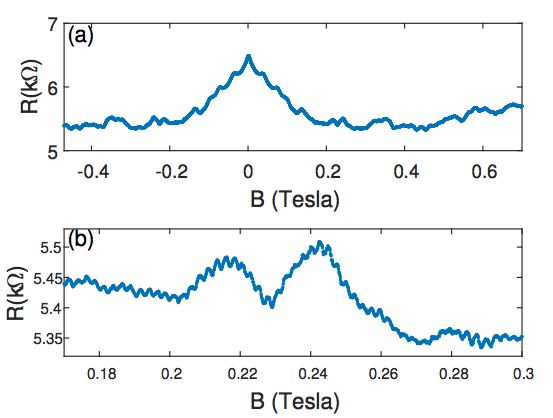}
\caption{(Color online) (a): The magnetoresistance of an Aharonov-Bohm ring at 1.5 K, as a function of the magnetic field, up to $B=$0.7 Tesla.  (b) The magnified data in the low-field region, showing the tiny oscillations superimposed on the Aharonov-Bohm ones. 
}
\label{Fig5}
\end{figure}

\begin{figure}
\includegraphics[width=9.2cm]{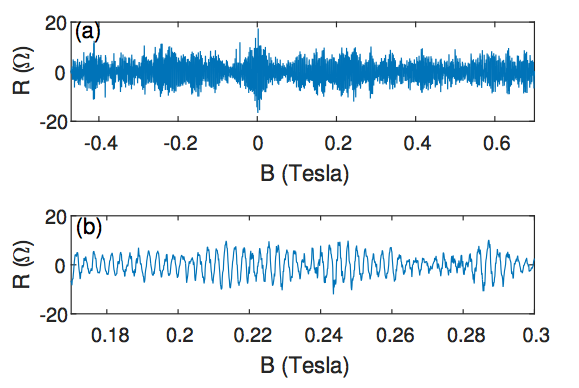}
\caption{(Color online) (a) and (b) The data shown in Figs. \ref{Fig5}(a) and \ref{Fig5}(b), 
once the low-frequency data points are filtered out.  
 }
\label{Fig6}
\end{figure}

\begin{figure}
\includegraphics[width=6cm]{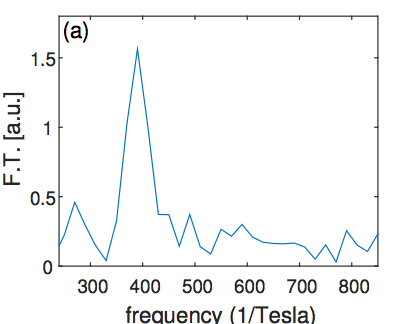}
\vspace*{0.5cm}
\includegraphics[width=6cm]{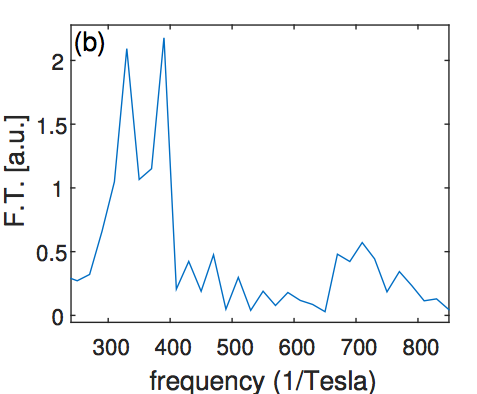}
\caption{(Color online) (a) The Fourier transform of the  magnetoresistance for magnetic fields in the  range  0.1$-$0.15 Tesla; the main peak is at $\approx$ 390 Tesla$^{-1}$.  (b) The Fourier transform of the  magnetoresistance for magnetic fields in the
range  0.65$-$0.7 Tesla,   where the peaks are at 
$\approx$ 330 Tesla$^{-1}$  and 
$\approx$
 390   Tesla$^{-1}$. }
\label{Fig7}
\end{figure}

The average area of the two rings we measured (see Sec. \ref{preparation} and Fig. \ref{Fig2})
is $\approx1.8\ \mu{\rm m}^{2}$; the periodicity of the AB oscillations with respect to the magnetic field is thus expected  to be $\approx 400$ Tesla$^{-1}$.
The magnetoresistance of  our ring A as a function of the magnetic field measured at   1.5 K   is portrayed in Fig. \ref{Fig5}. Panel (a) there depicts the raw data, and  panel (b)  magnifies the low-field  part of the data. 
Once the low-frequency data points are filtered out [see panels (a) and (b) in Fig. \ref{Fig6}], 
one can indeed observe fast oscillations with a frequency of about $400$ Tesla$^{-1}$, consistent with the estimated periodicity for the AB oscillations. On top of these, one sees beats, with a frequency of about $40$ Tesla$^{-1}$. These observations are consistent with the Fourier transform of the resistance, shown in Fig. \ref{Fig7}. Panel (a) there, (at magnetic fields in the range 0.1$-$0.15 Tesla)  is peaked around the expected AB frequency
$\approx 390$ Tesla$^{-1}$. Panel (b), based on data points from the range
0.65$-$0.7 Tesla,  has two peaks, at
$\approx 390$ Tesla$^{-1}$ and   at $\approx 330$ Tesla$^{-1}$. 
Analysis of data between these ranges shows a gradual decrease of the (average) AB frequency and a gradual increase of the splitting between the two peaks. Although the coherence length of our rings is of the order of the ring circumference (see below), Fig. \ref{Fig7}(b) also shows small peaks around $\approx 700$ Tesla$^{-1}$, probably corresponding to the second harmonic of the AB oscillations.

The splitting of the main peak in the power spectrum is the hallmark of the beating pattern,  \cite{Engel}
expected to result from the joint effect of the strong spin-orbit coupling and the Zeeman interaction. \cite{Aronov,Lyanda}
The appearance of the beating patterns, and their comparison with theoretical expectations, are discussed in Sec. \ref{beating}.

The Fourier transforms of the magnetoresistance of our sample B are similar to those shown in Fig. \ref{Fig7} for sample A.
The amplitude of the AB oscillations (the ``visibility"), and therefore also the heights of the leading peak in the Fourier transforms of the magnetoresistance, decrease with increasing temperature, because of the decrease of the coherence length. To deduce this length, we used measurements on our sample B, at magnetic fields below 0.05 Tesla, taken at 1.54 K,  1.78 K and 2.3 K.
The narrow range of magnetic fields  has been chosen because it contains mainly
an amplitude of only a ``single" harmonic.
According to Ref. \onlinecite{Milliken},  the amplitude of the $h/e$
oscillation in the conductance,  $\Delta G_{\rm AB}$, is
\begin{align}
\Delta G^{}_{\rm AB}=\frac{e^{2}}{h}\sqrt{\frac{\hbar D}{r^{2}k^{}_{\rm B}T}}\exp[-\pi r/L_{\phi}^{}]\ ,
\label{dGAB}
\end{align}
where $r$ is the radius of the ring and $D$ is the diffusion coefficient.


\subsection{The dephasing rate}
\label{rate}

\begin{figure}
\includegraphics[width=9cm]{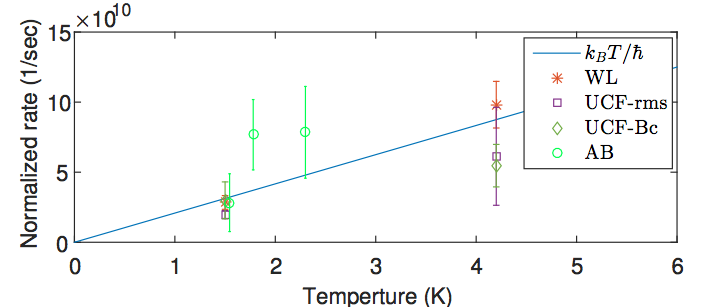}
\caption{(Color online) The ``normalized" dephasing rate extracted from all three experiments, as a function of the temperature. The solid line is $k_{\rm B}T/\hbar$, see Eq. (\ref{erate}).}
\label{Fig8}
\end{figure}

The dephasing rate $\tau^{-1}_{\phi}$ of electrons due to electron-electron interactions was calculated by Altshuler {\it et al.}; \cite{AKL} it is linearly proportional  
to the temperature and to the sheet resistance $R_{\square}$ of the sample, 
\begin{align}
\frac{1}{\tau^{}_{\phi}}
=\frac{k^{}_{\rm B}T}{\hbar}\frac{e^{2}R^{}_{\square}}{h}\ln\Big (\frac{h}{e^{2}R^{}_{\square}}\Big )\ .
\label{erate}
\end{align}
The dephasing rate is related to the coherence length by
\begin{align}
\tau^{-1}_{\phi}=D/L^{2}_{\phi}\ .
\end{align}
Using the known value of the two-dimensional density of states,  ${\cal N}(E)=0.2 \times 10^{11}$ cm$^{-2}$ meV$^{-1}$   for our InGaAs quantum well (with an effective mass of $m^{\ast}=0.05 m$),  and the Einstein relation $R_{\square} =1/[e^{2}{\cal N}(E)D]$,  together with the experimentally-determined value of dephasing length $L_{\phi}=3.8$ $\mu$m,   gives us self-consistency with   Eq. (\ref{erate}), with a diffusion coefficient of $D=0.13$ m$^{2}$/sec, and $R_{\square}=220$ $\Omega$ for the samples exploited in the WAL measurements. For the samples with the narrow etched mesa on which the UCF and the AB oscillations are measured, we estimate  
$D=0.06$ m$^{2}$/sec, and $R_{\square}=450$ $\Omega$.
The latter values are consistent with the change of the sheet resistance due to diffusive boundary scattering in two dimensions. \cite{Beenakker1}  
The symbols in Fig. \ref{Fig8}
mark the values of the ``normalized dephasing rate", $h/(\tau_{\phi}e^{2}R_{\square})\{\ln[h/e^{2}R_{\square})]\}^{-1}$ as extracted from our experiments. 
Ideally, these should fall on the straight line $k_{\rm B}T/\hbar$, and indeed, within the experimental error bars, they mostly  do, regardless of the coherence phenomenon from which they are deduced. In particular, at  $T=$1.5 K, where the experimental data are most reliable, 
all data points coincide with the theoretically-predicted value.

\section{The beating patterns in the magnetoconductance of the rings}
\label{beating}

The combined effect of strong spin-orbit  and 
Zeeman interactions, in the adiabatic limit,  is expected to induce a Berry phase on the spin part of the electronic wave function. The possibility that this geometrical phase 
can be detected in power spectra of the magnetoconductance oscillations of mesoscopic rings 
has been pursued 
quite actively, both theoretically and experimentally (see Sec. \ref{intro} for a brief survey). An interesting (theoretical) observation has been made in Ref. \onlinecite{Engel}. Carrying out numerically a rather complicate calculation 
of the AB oscillations and the corresponding power spectrum (computed by zero-padding the data before applying the Fourier transform code), the authors found that  the peak splitting in diffusive rings
depends strongly on the 
different dephasing sources, and that for small dephasing  the splitting is totally masked.

\begin{figure}
\vspace*{0.5cm}
\includegraphics[width=7.5cm]{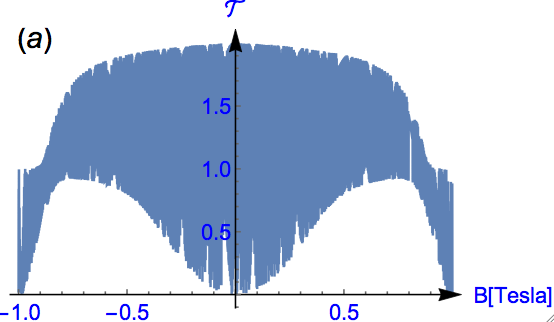}
\vspace{1.75cm}
\includegraphics[width=7.5cm]{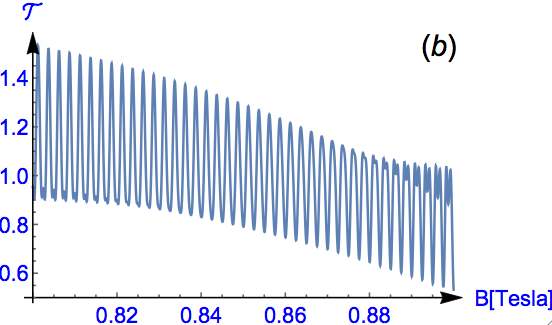}
\caption{(Color online)  The transmission, Eq. (\ref{tau}), as a function of the magnetic field over a wider range of fields (a) and over a restricted range (b). The parameters are given in Secs. \ref{intro} and \ref{beating}. }
\label{Fig9}
\end{figure}

Our data are not sufficient to examine this observation. We have therefore analyzed the simpler expression given in Ref. \onlinecite{Aronov} for the transmission ${\cal T}$ of a clean Aharonov-Bohm ring \cite{com} subjected to strong spin-orbit and Zeeman interactions, 
\begin{align}
{\cal T}=\Big [1+\frac{{\rm tan}^{2}(\Phi^{+}_{t}/2)}{4\sin^{2}(\Phi^{+}_{S}/2)}\Big ]^{-1}+
\Big [1+\frac{{\rm tan}^{2}(\Phi^{-}_{t}/2)}{4\sin^{2}(\Phi^{-}_{S}/2)}\Big ]^{-1}\ .
\label{tau}
\end{align}
This expression is valid in the adiabatic limit, pertaining to the case where, as mentioned in Sec. \ref{intro},  both $\omega_{\rm so}$ and $\omega_{\rm Z}$ are larger than the rotation frequency around the ring, $\Omega$. \cite{Aronov} This condition is  fulfilled by our rings, whether the rotation frequency  is calculated in the clean limit, $\Omega=  v_{\rm F}/(2\pi r)$, leading to $\hbar\Omega\approx 
0.33$ meV, or in the diffusive limit, $\Omega =D/[(2\pi r)^{2}]$, in which case 
$\hbar\Omega\approx 
0.001$ meV. 
The transmission ${\cal T} $ is given in terms of   two phases, each of which is  different   for the two spin orientations. The phase $\Phi^{\pm}_{ t}$ comprises the Aharonov-Bohm phase and the Berry phase, $\Phi_{\rm B}$, 
\begin{align}
\Phi^{\pm}_{t}=\Phi^{\pm}_{\rm B}-2\pi\frac{\Phi}{\Phi^{}_{0}}\ ,\  \Phi^{\pm}_{\rm B}=
\pm \pi\Big (1-\frac{\omega^{}_{\rm Z}}{\sqrt{\omega^{2}_{\rm so}+\omega^{2}_{\rm Z}}}\Big )\ , 
\label{phit}
\end{align}
where $\Phi$ is the magnetic flux through the ring. With our experimental parameters, $\Phi/\Phi_{0}\approx
400 \times B$, where $B$ is measured  Tesla. The other phase, $\Phi^{\pm}_{S}$ (termed ``standard" in Ref. \onlinecite{Aronov}), is in fact the optical path along the ring perimeter; it is different for each spin direction since the Zeeman energy modifies the Fermi energy of each spin. This phase is given by
\begin{align}
\Phi^{\pm}_{S}=2\pi r k^{\pm}_{0}\ ,
\label{phis}
\end{align}
where $k^{\pm}_{0}$ are the solutions of 
\begin{align}
E^{}_{\rm F}=\frac{(\hbar k^{\pm}_{0})^{2}}{2m^{\ast}}\pm\hbar
\sqrt{\omega^{2}_{\rm Z}+(\omega^{}_{\rm so}k^{\pm}_{0}/k^{}_{\rm F})^{2}}\ .
\label{ef}
\end{align}
The effective electron mass $m^{\ast}$ in our samples is $\approx 0.05$ times the free-electron mass, and the Fermi energy $E_{\rm F}\approx 19.6$ meV.  Solving Eq. (\ref{ef}) yields
\begin{align}
[k^{\pm}_{0}&]^{2}=k^{2}_{\rm F}\Big [1+\frac{1}{2}\Big (\frac{\hbar\omega^{}_{\rm so}}{E^{}_{\rm F}}\Big )^{2}\Big ]\nonumber\\
&\mp\frac{1}{E^{}_{\rm F}}\sqrt{(\hbar\omega^{}_{\rm Z})^{2}+(\hbar\omega^{}_{\rm so})^{2}\Big [1+
\frac{1}{4}\Big (\frac{\hbar\omega^{}_{\rm so}}{E^{}_{\rm F}}\Big )^{2}\Big ]}\ .
\label{kpm}
\end{align}
For our samples' parameters $E_{\rm F}\approx 12.3\times \hbar\omega_{\rm so}$, while $\omega_{\rm Z}$ becomes comparable to $\omega_{\rm so}$ at about $B=2$ Tesla. The transmission as a function of the magnetic field as derived from Eq. (\ref{tau}) is illustrated in Figs. \ref{Fig9} (the parameters use are those quoted in Sec. \ref{intro}  and above).

The two panels in Fig. 9 display the transmission for two different ranges of the magnetic  field. Both show an envelope of the AB oscillations, which varies slowly. Figure \ref{Fig9}(a)
clearly exhibits beats, superimposed on fast AB oscillations. From Eq. (\ref{phit}), the Berry phase is of order $\pi$, and the AB phase is of order $2\pi\times 
400\times B$ ($B$ in Tesla). Therefore, the Berry phase affects the results only for $B<0.002$ Tesla, and it is practically irrelevant for the interpretation of our data.
Equation (\ref{tau}) shows that the modulations of the  AB oscillations, which result from the factors $\tan^2(\Phi^\pm_t/2)\approx\tan^2(\pi\Phi/\Phi^{}_0)$, are modified by the prefactors $\sin^2(\Phi^\pm_S/2)$, which create beats due to the dependence of $\Phi^\pm_S$ on $B$.
At $\hbar\omega^{}_{\rm so}=0$, Eqs. (\ref{phis})-(\ref{kpm}) yield $\Phi^{\pm}_{S}/(2\pi)=118\times\sqrt{1\mp \hbar\omega^{}_Z/E^{}_{\rm F}}\approx 118 \mp 2.7 \times B$ ($B$ in Tesla). Then the transmission given in Eq. (\ref{tau}) 
should exhibit  beats at a very small frequency of order 2.7 Tesla$^{-1}$. In our samples $\hbar\omega^{}_{\rm so}\approx 1.6$ meV, and then the two functions $\Phi^{\pm}_{S}$ are approximately parabolic in $B$ (for small $B$), with slopes that increase with $B$. Specifically, one has $\Phi^{+}_{S}/(2\pi)\approx 114-0.14 (B-0.1)+O[(B-0.1)^2]$ and $\Phi^{-}_{S}/(2\pi)\approx 123+0.14 (B-0.1)+O[(B-0.1)^2]$ near $B=0.1$ Tesla, while $\Phi^{+}_{S}/(2\pi)\approx 113-0.98 (B-0.7)+O[(B-0.7)^2]$ and $\Phi^{-}_{S}/(2\pi)\approx 124+0.89 (B-0.7)+O[(B-0.7)^2]$ near $B=0.7$ Tesla. The corresponding beats have even smaller frequencies, of order 0.14 Tesla$^{-1}$ and 1 Tesla$^{-1}$, respectively. These frequencies seem consistent with the envelopes of the fast oscillations in Fig. \ref{Fig9}. Although the theory exhibits a slow decrease of the average frequency, and a gradual increase of the beating frequencies, similar to the experimental observations, all of these theoretical beat frequencies are  much smaller than those seen in the experiments. Fourier transforms of the data in Fig. \ref{Fig9} (with or without zero-padding) indeed yield single peaks at the first harmonic of the AB oscillations, somewhat broadened by the Zeeman contributions. Higher harmonics do show small splittings of the peaks.

\section{Summary}
\label{summary}

We have measured weak antilocalization effects, universal conductance fluctuations,  and Aharonov-Bohm oscillations in the
two-dimensional electron gas formed  in  InGaAs/AlInAs heterostructures. This system  possesses strong spin-orbit coupling and a high Land\'{e} factor.
Phase-coherence
lengths of  2$-$4 $\mu$m at 1.5$-$4.2 K were extracted from the magnetoconductance measurements.
The analysis of the coherence-sensitive data reveals that the temperature dependence of the decoherence  rate complies with the dephasing mechanism originating from electron-electron interactions in all three experiments. 

Distinct beating patterns superimposed on  the Aharonov-Bohm oscillations are observed over a wide range of magnetic fields,  up to 0.7 Tesla at the relatively high temperature of 1.5 K. The Berry phase is much smaller than the AB phase, and therefore cannot be responsible for these beats. Qualitatively, the theory of Aronov and Lyanda-Geller\cite{Aronov} does exhibit beats due to the interplay between the Zeeman and the spin-orbit interactions. However, the beating frequencies found in this theory are much smaller than those observed experimentally.
It thus seems that  the source of the beating pattern in the magnetoconductance of our rings are the different  electronic paths through the ring, each penetrated  by  a slightly different magnetic flux. \cite{Aharony} For example, since the AB frequencies are proportional to the area encompassed by the electronic paths, the measured ratio of the two frequencies in Fig. \ref{Fig7}(b), i.e., $390/330\approx 1.2$,  implies a radii ratio of about 1.1.  The width of our rings (see Fig. \ref{Fig2}) can easily accommodate two paths with such a radii ratio, and hence may explain the beating pattern.

\begin{acknowledgments} We thank Y. Lyanda-Geller for very useful comments. This work was partially supported by the Israeli Science Foundation
(ISF) grant 532/12 and grant 252/11, and by the infrastructure program of Israel
Ministry of Science and Technology under contract
3-11173.
\end{acknowledgments}


\begin{thebibliography}{99}

\bibitem{Fert}
A. Fert, Rev. Mod. Phys. {\bf 80}, 1517 (2008).


\bibitem{Wolf}
 S. A. Wolf,  D. D. Awschalom, R. A. Buhrman, J. M. Daughton, S. von Moln\'{a}r, M. L. Roukes, A. Y. Chtchelkanova, D. M. Treger, Science {\bf 294}, 1488 (2001).
 
 \bibitem{Zutic}
 I.  \u{Z}uti\'{c}, J. Fabian, and S. Das Sarma, Rev. Mod. Phys.
{\bf 76}, 323 (2004). 
 
 
 \bibitem{Bratkovsky}
A. M. Bratkovsky, Reports on Progress in Physics {\bf 71},
026502 (2008).
 
\bibitem{Altshuler}
B. L. Altshuler, A. G. Aronov, and D. E. Khmel'nitskii, J.  Phys. C {\bf 15}, 7367 (1982).

\bibitem{Hikami}
S. Hikami, A. I. Larkin, and Y. Nagaoka, Progress of Theoretical Physics {\bf 63}, 707 (1980).


\bibitem{Maekawa}
S. Maekawa and H. Fukuyama, 
J. Phys. Soc. Jpn. {\bf 50}, 2516 (1981).


\bibitem{Lee}
P. A. Lee, A. D. Stone, and H. Fukuyama, Phys. Rev. B
{\bf 35}, 1039 (1987).




\bibitem{Aharonov}
Y. Aharonov and D. Bohm, Phys. Rev. {\bf 115}, 485 (1959).

\bibitem{Washburn}
S. Washburn and R. A. Webb, Advances in Physics {\bf 35},
375 (1986).

\bibitem{Milliken}
P. Milliken, S. Washburn, C. P. Umbach, R. B. Laibowitz, and R. A. Webb, Phys. Rev. B {\bf 36}, 4465 (1987).

\bibitem{Rashba}
 E. I. Rashba, Fiz. Tverd. Tela (Leningrad) {\bf 2}, 1224 (1960) [Sov. Phys. Solid State {\bf 2}, 1109 (1960)]; Y. A. Bychkov and E. I. Rashba, J. Phys. C {\bf 17}, 6039 (1984).

\bibitem{Winkler}
 R. Winkler, {\it Spin-Orbit Coupling Effects in Two-Dimensional Electron and Hole Systems} (Springer-Verlag, Berlin, 2003).

\bibitem{Schapres}
Th. Sch\"{a}pres, E. Engels, J. Lange, Th. Klocke, M. Hollfelder, and H. L\"{u}th, J. Appl. Phys. {\bf 83}, 4324 (1998).


\bibitem{Nitta}
S. Nitta, H. Choi, and S. Yamada, Physica E  
{\bf 42}, 987 (2010).

\bibitem{Aronov}
A. G. Aronov and Y. B. Lyanda-Geller, Phys. Rev. Lett.
{\bf 70}, 343 (1993); see also   
F. E. Meijer, A. F. Morpurgo, and T. M. Klapwijk, 
Phys. Rev. B {\bf 66}, 033107 (2002).





\bibitem{Berry}
M. V. Berry, Proc. Royal Soc. A {\bf 392}, 45 (1984).




\bibitem{Engel}
D. Loss, P. Goldbart, and A. V. Balatsky, Phys. Rev. Lett. {\bf 65}, 1655 (1990);  H.-A.	Engel and D. Loss, Phys. Rev. B {\bf 62}, 10238 (2000). 


\bibitem{Stern}
A. Stern, Phys. Rev. Lett. {\bf 68}, 1022,  (1992).

\bibitem{vanLangen}
S. A. van Langen, H. P. A. Knops, J. C. J. Paasschens, and C. W. J. Beenakker, 
Phys. Rev. B {\bf 59}, 2102 (1999).

\bibitem{Loss}
D. Loss, H. Schoeller, and P.M. Goldbart, Phys. Rev. B {\bf 48}, 15218 (1993);
Phys. Rev. B {\bf 59}, 13328 (1999).





\bibitem{Morpurgo}
A. F. Morpurgo, J. P. Heida, T. M. Klapwijk, B. J. van
Wees, and G. Borghs, 
Phys. Rev. Lett. {\bf 80}, 1050 (1998).



\bibitem{Yau}
J-B. Yau, E. P. De Poortere, and M. Shayegan, Phys. Rev. Lett. {\bf 88}, 
146801 (2002).




\bibitem{Lyanda}
M. J. 
Yang,   C. H. Yang, and Y. B. Lyanda-Geller,  
Europhys. Lett.   {\bf 66},  826 (2004).






\bibitem{Grbic}
B. Grbi\'{c}, R. Leturcq, T. Ihn, K. Ensslin, D. Reuter, and
A. D. Wieck, 
Phys. Rev. Lett. {\bf 99}, 176803 (2007). 


\bibitem{Meijer}
F. E. Meijer, A. F. Morpurgo, T. M. Klapwijk, T. Koga,
and J. Nitta, 
Phys. Rev. B {\bf 69}, 035308 (2004).



\bibitem{Ren}
S. L. Ren, J. J. Heremans, C. K. Gaspe, S. Vijeyaragunathan, T. D. Mishima, and M. B. Santos, J. Phys.: Condens. Matter {\bf 25}, 435301 (2013).

\bibitem{Aharony}
A. Aharony, O. Entin-Wohlman, T. Otsuka, S. Katsumoto, H. Aikawa, and K. Kobayashi, Phys. Rev. B {\bf 73}, 195329 (2006).




\bibitem{Zeng}
Y. Zeng, X. Cao, L. Cui, M. Kong, L. Pan, B. Wang, and
Z. Zhu, J.of Crystal Growth {\bf 210},  227228 (2001).




\bibitem{Hoke}

W. E. Hoke, T. D. Kennedy, A. Torabi, C. S. Whelan, P. F. Marsh, R. E. Leoni, C. Xu, and K. C. Hsieh, J. 
Crystal Growth {\bf 251}, 827 (2003).




\bibitem{Iordanskii}
S. V. Iordanskii, Yu. B. Lyanda-Geller, and G. E. Pikus, 
PisÕma
Zh. Eksp. Teor. Fiz. {\bf 60}, 199 (1994) [JETP Lett. {\bf 60}, 206 (1994).

\bibitem{Knap}
W. Knap, C. Skierbiszewski, A. Zduniak, E. Litwin-Staszewska, D. Bertho, F. Kobbi, J. L. Robert, G. E. Pikus, F. G. Pikus, S. V. Iordanskii, V. Mosser, K. Zekentes, and Yu. B. Lyanda-Geller, 
Phys. Rev. B {\bf 53}, 3912 (1996).



\bibitem{Beenakker}
C. W. J. Beenakker and H. van Houten, Solid State Physics {\bf 44}, 1 (1991).


\bibitem{Chandrasekhar}
V. Chandrasekhar, P. Santhanam, and D. E. Prober, Phys. Rev. B {\bf 42}, 6823 (1990).

\bibitem{AKL}
B. L. Altshuler, D. Khmel'nitskii, A. I. Larkin, and P.  A. Lee, Phys. Rev. B. {\bf 22}, 5142 (1980).


\bibitem{Beenakker1}
C. W. J. Beenakker and H. van Houten, Phys. Rev. B  {\bf 38}, 3232 (1988).



\bibitem{com}
Reference \onlinecite{Lyanda} gives another expression for the transmission. However, the setup considered in that paper is different from ours.
 
\end{thebibliography}
\end{document}